\documentclass[12pt]{article}

\usepackage{amsmath}
\usepackage{latexsym}
\newcommand\be{\begin{equation}}
\newcommand\ee{\end{equation}}
\newcommand\bea{\begin{eqnarray}}
\newcommand\eea{\end{eqnarray}}
\newcommand\beas{\begin{eqnarray*}}
\newcommand\eeas{\end{eqnarray*}}

\def\tr{{\rm Tr}}

\begin{document}
%\right{WITS-CTP-029}
\title{``Non-renormalization" without supersymmetry} 
\author{\\
Jo\~ao P. Rodrigues\footnote{Email: rodriguesj@physics.wits.ac.za} \\
\\
School of Physics and Centre for Theoretical Physics \\
University of the Witwatersrand\\
Johannesburg, Wits 2050 \\ South Africa \footnote{WITS-CTP-029}\\
}

%\begin{document}
\maketitle

\begin{abstract}
The $g_{YM}$ perturbed, non supersymmetric extension of the dual single matrix description 
of $1/2$ BPS states,  
within the Hilbert space reduction to the oscillator subsector associated with chiral primaries
is considered.
This matrix model is described in terms of a single hermitean matrix. 
It is found that, apart from a trivial shift in the energy, 
the large $N$ background, spectrum and interaction
of invariant states \it{are independent} \rm of $g_{YM}$. 
This property applies to more general D terms.
\end{abstract}  

\newpage

\noindent
\section{Background and Motivation}

\noindent
Recent studies of giant gravitons in AdS backgrounds [1-8] have led to a  
dual matrix model description of 1/2 BPS states in terms of the large N limit of a single complex matrix 
\cite{Corley:2001zk} in a 
harmonic potential \cite{Berenstein:2004kk}.

%One may motivate for this 
%We briefly review the precise identification of variables resulting from the dual description 
%of 1/2 BPS states \cite{Corley:2001zk}. 
This can be motivated 
%arrived at 
as follows:
starting with the leading Kaluza Klein compactification of the bosonic
sector of $\cal{N}=$ $4$ SYM on $S^3 \times R$, one chooses the plane defined by two adjoint scalars 
($N\times N$ hermitean matrices) $X_1$ and $X_2$. 
%grouped into a complex matrix $Z=X_1+i X_2$. 
The corresponding bosonic sector of the Hamiltonian is

\be\label{HamX}
\hat{H}
\equiv
\frac{1}{2} \tr (P_1^2 ) + \frac{1}{2} \tr (P_2^2 ) + \frac{w^2}{2} \tr (X_1^2) + \frac{w^2}{2} \tr (X_2^2) 
- g_{YM}^2 \tr [X_1,X_2][X_1,X_2]
\ee

\noindent
with $P_1$ ($P_2$) canonical conjugate to $X_1$ ($X_2$, respectively). The harmonic potential arises as a result of the
coupling to the curvature of the manifold.

In terms of the complex matrix $Z=X_1+i X_2$ and canonical momentum $\Pi = 1/2 (P_1 - i P_2) $,

\be\label{HamZ}
\hat{H}
=
2 \tr \Pi^{\dagger} \Pi + \frac{w^2}{2} \tr (Z^{\dagger} Z) +
\frac{g_{YM}^2}{4} \tr [Z,Z^{\dagger}][Z,Z^{\dagger}].
\ee

\noindent
Introducing matrix valued creation and annhilation operators 

$$
      Z=\frac{1}{\sqrt{w}}  (A + B^{\dagger}) \qquad \Pi =-i\frac{\sqrt{w}}{2} ( A^{\dagger} - B),
$$

\noindent
motion in this plane is then characterized by the energy $E$ and the two dimensional angular momentum:

\be\label{Ei}
                \hat{J}=    \tr (A^{\dagger} A) -  \tr (B^{\dagger} B) .
\ee

\noindent
A(and B) quanta carry well defined charge $1$ ($-1$, respectively).

%\noindent
When imbedded in $\cal{N}$$=4$ SYM, the $g_{YM}$ dependent interaction in (\ref{HamZ}) is one of the 
so called D terms, and is subject to non-renormalization theorems. One may then consider the free theory.
In this case, the (invariant) eigenstates are $\tr((A^{\dagger})^n)\tr((B^{\dagger})^m) |0>$.

1/2 BPS states then correspond to a restriction of the theory to the 
(chiral) sector with no $B$ excitations, for which $E=E_0= wJ$ ($E_0$ is the free theory energy)
{\footnote{The B sector can also be projected out by taking a pp wave limit \cite{Berenstein:2002jq}. This requires
$J$ to become large.}}.

%\noindent
When restricted to correlators of these chiral primary operators, the dynamics of the system is fully described 
by free fermions in the harmonic oscillator potential. It actually turns out, as shown in \cite{Lin:2004nb}, that the 
gravity description of 1/2 BPS states is completely determined by a phase space density function 
associated with a general fermionic droplet configuration. 

%\noindent
As suggested in \cite{Donos:2005vm}, the dynamics of the $A$, $A^{\dagger}$ system can be described in terms of an
hermitean matrix $M$ 
\be\label{Ide}
       M \equiv \frac{1}{\sqrt{2w}}( A + A^{\dagger})  \quad P_ M = -i \sqrt{\frac{w}{2}}( A - A^{\dagger}), 
\ee

\noindent
while retaining $B$, $B^{\dagger}$ creation and annhilation operators. The change of variables is:

\bea\label{CVar}
X_1 = \frac{1}{\sqrt{2}} M + \frac{1}{2\sqrt{w}}(B+B^{\dagger}) &\quad& 
X_2 = \frac{1}{\sqrt{2}w} P_M + \frac{i}{2\sqrt{w}}(B-B^{\dagger}) \\
P_1 = \frac{1}{\sqrt{2}} P_M - \frac{i\sqrt{w}}{2}(B-B^{\dagger})&\quad& 
P_2 = - \frac{w}{\sqrt{2}} M + \frac{\sqrt{w}}{2}(B+B^{\dagger}) \nonumber
\eea

\noindent
As shown in \cite{Donos:2005vm}, in the 1/2 BPS sector, the gravitational degrees of freedom emerge in a natural way
from the density of eigenvalues description of the large $N$ dynamics of the hermitean matrix $M$. The energy and flux 
of the 1/2 BPS states obtained in \cite{Lin:2004nb} are exactly those of the leading large $N$ configuration 
of the hermitean matrix $M$ in a harmonic potential, in a bosonic phase space density description
\footnote{This is already suggested in (\ref{CVar}), as $X_1 = 1/{\sqrt{2}} M + ...$ and 
$X_2 = {1}/{\sqrt{2}w} P_M + ...$}. This will be briefly reviewed in the next section. Fluctuations about the large $N$ matrix configuration were shown to
agree with fluctuations \cite{Donos:2005vm}, \cite{Grant:2005qc} about the gravity background of \cite{Lin:2004nb}.    

%\noindent
The purpose of this note is to investigate, in the absence of supersymmetry, the non-perturbative consequences of
the quartic $g_{YM}$ interaction in (\ref{HamX}) while still restricting the theory to the sector with no $B$ excitations.
In other words, the system of interest is 

\be\label{HamAA}
\hat{H}
=
w \tr (A^{\dagger} A)  +
\frac{g_{YM}^2}{4w^2} \tr [A^{\dagger},A][A^{\dagger},A].
\ee

In order to study its large $N$ limit, 
I will use the variables (\ref{CVar}), in terms of which the hamiltonian (\ref{HamAA}) becomes:

\be\label{HamA}
\hat{H_{A}}
= \frac{1}{2} \tr P_M^2 + \frac{w^2}{2} \tr M^2 -
\frac{g_{YM}^2}{4 w^2} \tr [M,P_M][M,P_M].
\ee

\noindent
In the absence of supersymmetry, the argument for the decoupling of $B$ excitations is potentially weakened. It should
always be a good approximation for external states with large $J$ charges \cite{Weinberg:1966jm} and for large $w$. 
However, the hamiltonian (\ref{HamAA}) (or (\ref{HamA})) is of great interest \it{per se} \rm
%in its own merit, 
as it contains an interaction with the structure typical of a Yang-Mills interaction.

Remarkably, we will find that, apart from a trivial shift in the energy, the large $N$ background, spectrum and interaction
of gauge invariant states \it{are independent} \rm of $g_{YM}$, \it{even in the absence of supersymmetry}\rm .

%\noindent
This letter is organized as follows: within the collective field theory approach \cite{Jevicki:1979mb}, a general 
argument is presented in Section $2$ for the ``non-renormalization' properties of the theory, 
and an explicit non-perturbative argument is developed in Section $3$. A simple diagrammatic check is
carried out Section $4$, and a generalization presented in Section $5$. Section $6$ is reserved for a brief conclusion.

\noindent
\section{A general argument}
In order to obtain the large $N$ limit of (\ref{HamA}), we will make use of collective field theory approach 
\cite{Jevicki:1979mb}. The starting point of this approach is to consider the action of the Hamiltonian 
on wave functionals of gauge invariant
operators, i.e., operators invariant under the transformation:

\be\label{transf}
M \to U^{\dagger} M U , \qquad U \, \, \rm{unitary} .
\ee

For a single matrix model such as (\ref{HamA}), these can be chosen as the density of eigenvalues 
$\lambda_i , i=1,...,N$ of the matrix $M$:

\be
        \phi(x) = \int \frac{dk}{2\pi} e^{-ikx} Tr(e^{ikM}) = \sum_{i=1}^N \delta(x-\lambda_i),
\ee

\noindent
or its fourier tranform

\be
        \phi_k = Tr(e^{ikM})
\ee

\noindent
Let us analyse the structure of the $g_{YM}$ dependent operator in (\ref{HamA}) in more detail. It can be written as

\be\label{gauss}
-\frac{g_{YM}^2}{4 w^2} \tr [M,P_M][M,P_M] = -\frac{g_{YM}^2}{4 w^2} \hat{G}_{ij}\hat{G}_{ji} + \frac{g_{YM}^2 N^3}{4 w^2}
\ee

\noindent
where 

\be
                                 \hat{G}_{ij} \equiv M_{ik} \hat{P}_{kj} - M_{kj} \hat{P}_{ik}.
\ee

\noindent
We recognize the generators of the transformation (\ref{transf}). Therefore, when restricted to the gauge invariant subspace,
the $g_{YM}$ term in (\ref{HamA}) does not contribute, except for the trivial constant shift in energy in (\ref{gauss})
\footnote{One may also think of the system as gauged, in which case the invariance under (\ref{transf})
results from Gauss' law, and restriction to wave functionals of gauge invariant operators explicitly satisfies Gauss'
law}.

\noindent
\section{Explicit non-perturbative argument}

Because, as a result of the use of the chain rule in the kinetic energy operator, the interactions in the theory
organize themselves with different powers of $N$, it is important to provide an explicit verification of the general
argument presented in the previous section.

%\noindent
What is diferent from previous applications of the collective field theory to the large $N$ limit of the single matrix
hamiltonian (\ref{HamA}), is the sigma model nature of the kinetic energy operator and the presence of terms 
linear in momentum:

\bea\label{Kinetic}
&&T = \frac{1}{2} g_{i_1,i_2,i_3,i_4}(M) P_{i_1,i_2} P_{i_3,i_4} - \frac{i}{2} s_{i_1,i_2}(M) P_{i_1,i_2}\nonumber \\
&&\equiv \frac{1}{2} g_{AB}(X)P_A P_B - \frac{i}{2} s_A(X)P_A, \quad  
\quad P_{i_1,i_2}\equiv - i \frac{\partial}{\partial{M}_{i_2,i_1}} \nonumber \\
&&g_{i_1,i_2,i_3,i_4}(M) = \delta_{i_1,i_4} \delta_{i_2,i_3} 
+\frac{g_{YM}^2}{w^2}\left((M^2)_{i_4,i_1}\delta_{i_2,i_3}- M_{i_4,i_1} M_{i_2,i_3} \right) ,\nonumber \\
&&s_{i_1,i_2}(M)= - \frac{g_{YM}^2}{w^2} \left( N M_{i_2,i_1} - \tr(M) \delta_{i_1,i_2} \right) .
\eea

\noindent
Denoting by $\phi_{\alpha}$ a generic gauge invariant variable, 
the kinetic energy operator (\ref{Kinetic}) takes the form, when acting on functionals of $\phi_{\alpha}$,

\bea
&&T = - \frac{1}{2} g_{AB}(X) \frac{\partial}{\partial X_A} \frac{\partial}{\partial X_B} 
- \frac{1}{2} s_{A}(X) \frac{\partial}{\partial X_A} =
- \frac{1}{2} \left( \bar{\omega}_{\alpha} \frac{\partial}{\partial {\phi_{\alpha}}}
+ \Omega_{\alpha,\beta} \frac{\partial}{\partial {\phi_{\alpha}}}  \frac{\partial}{\partial {\phi_{\beta}}} \right)
\nonumber \\
&&\bar{\omega}_{\alpha} = g_{AB}(X) \frac{\partial^2 \phi_{\alpha}}{\partial X_A \partial X_B} 
+ s_{A}(X) \frac{\partial \phi_{\alpha}}{\partial X_A}
\quad
\Omega_{\alpha,\beta} = g_{AB}(X) \frac{\partial\phi_{\alpha}}{\partial X_A} \frac{\partial\phi_{\beta}}{\partial X_B}
\eea

\noindent
For $\Omega_{\alpha,\beta}$, I obtain

\be\label{BigOm}
\Omega_{k,k'} = - k k' \phi_{k+k'}
\ee

\noindent
Due to the antisymmetric nature of the Yang-Mills interaction reflected in the sigma model nature of the kinetic energy
(\ref{Kinetic}), the result above for $\Omega_{k,k'}$ is independent of $g_{YM}$ and is the same as that for the standard
kinetic energy. 
%This result is at the root of the large $N$ solvability of the system (\ref{HamA}).

%\noindent
For $\bar{\omega}_{\alpha}$, I obtain  

\bea\label{LittleOm}
&&\bar{\omega} (x) = - 2 \partial_x \times \\
&&\Big{\{} \phi(x) \left[ \int dy \frac{\phi(y)}{x-y} 
+ \frac{g_{YM}^2}{2 w^2}\int dy (x-y) \phi(y) 
+ \frac{g_{YM}^2}{2 w^2}\left(\int dy y \phi(y) - N x \right) \right]\Big{\}} \nonumber \\
&&= - 2 \partial_x \Big{\{} \phi(x) \left[ \int dy \frac{\phi(y)}{x-y}
+ \frac{g_{YM}^2}{2 w^2} x \left(\int dy \phi(y) - N \right) 
\right]\Big{\}} \nonumber
\eea

\noindent
Interpreting $N=\tr (1) = \int dx \phi(x)$, one would immediatly conclude that both (\ref{BigOm}) and (\ref{LittleOm}) 
are independent of $g_{YM}$, proving our result. 
However, we follow the more rigorous approach of imposing this constraint via a Lagrange multiplier, and
choose to enforce the constraint after variation.

%\noindent
Due to the change of variables $X_A \to \phi_{\alpha}$, one performs the similarity transformation 
\cite{Jevicki:1979mb} 
induced by the Jacobian $J$:

\be
\frac{\partial}{\partial \phi_{\alpha}}
\to J^{\frac{1}{2}} \frac{\partial}{\partial \phi_{\alpha}} J^{-\frac{1}{2}}  \qquad
\Omega_{\alpha,\beta} \frac {\partial \ln J}{\partial\phi_{\beta} } = \omega_{\alpha},
\ee

\noindent
where only the leading (in $N$) expression for $\ln J$ is described. One obtains the form of the 
collective field hamiltonian sufficient for the description of the leading large $N$ background and fluctuations:

\bea\label{CollBef}
\hat H_{eff} &=&
\frac{1}{2} \frac{\partial}{\partial \phi_{\alpha}} \Omega_{\alpha,\beta} \frac{\partial}{\partial \phi_{\beta}}
+ \frac{1}{8} \bar{\omega}_{\alpha} {\Omega}^{-1}_{\alpha,\beta} \bar{\omega}_{\beta} 
+ \int dx \frac{1}{2} w^2 x^2 \phi(x) \nonumber \\
&-& \mu \left(\int dx \phi(x) - N) \right) = \nonumber \\ &-& \frac{1}{2}
\int dx \partial_x {\partial \over \partial \phi(x)} \phi(x) \partial_x {\partial \over \partial \phi(x)}
\nonumber \\
&+& \frac{1}{2} \int dx \phi(x) \left[ \int \frac{dy \phi(y)}{x-y}
+  \frac{g_{YM}^2}{2 w^2} x (\int dy \phi(y) - N) \right]^2
\nonumber \\
&+& \int dx \frac{1}{2} w^2 x^2 \phi(x) 
- \mu \left( \int dx \phi(x) - N) \right). 
\eea

\noindent
The lagrange multiplier $\mu$ enforces the contraint

\be\label{constraint}
\int dx \phi(x) = N.
\ee

\noindent
To exhibit explicitly the $N$ dependence, we rescale

\bea \label{Rescaling}
x  \to \sqrt{N} x \, , &\quad& \psi(x,0) \to \sqrt{N} \psi(x,0) \, , \\
\mu \to N \mu \, , &\quad& 
-i {\partial \over \partial \psi(x,0)}\equiv \Pi(x) \to {1 \over N} \Pi(x). \nonumber
\eea

\noindent
Using the identity 

\be
\int dx \phi(x) \left[ \int \frac{dy \phi(y)}{x-y}\right]^2 = \frac{\pi^2}{3} \int dx \phi^3(x),
\ee

\noindent
we obtain

\bea \label{HEff}
H_{eff}&=& {{1 \over 2N^2}}
\int dx \partial_x \Pi(x) \phi(x) \partial_x \Pi(x) \nonumber \\ 
&+& N^2 \Big[  \int dx {\pi^2 \over 6}\phi^3(x) +   
\frac{\lambda^2}{8w^4} \big( \int dx x^2 \phi(x) \big) \big( \int dy \phi(y) - 1 \big)^2 \nonumber \\
&+&\frac{\lambda}{4w^2} \big( \int dx \phi(x) \big)^2\big( \int dy \phi(y) - 1 \big) +
\nonumber \\
&+& \int dx \phi(x){w^2 x^2 \over 2}- \mu \big( \int dx \phi(x) - 1 \big) \Big], 
\eea

\noindent
where $\lambda=g_{YM}^2 N$ is 't Hooft's coupling.

%\noindent
%Comment on multi trace terms in other approaches

%\noindent
%\section{Large $N$ background}

%\noindent
The large $N$ configuration is the semiclassical background corresponding to 
the minimum of the effective potential in (\ref{HEff}), and satisfies

\be\label{PiPhi}
\frac{\pi^2}{2} \phi_0^2 + \frac{\lambda}{4w^2}  
+ \frac{1}{2} w^2 x^2 - \mu = 0,
\ee 

\noindent
where the constraint $\int dx \phi_0(x)=1$ has been applied after variation. This constraint fixes 
$\mu= w + \frac{\lambda}{4w^2}$, and we arrive at the Wigner distribution background

\be\label{NewWign}
\pi \phi_0 (x) = \sqrt{2 w -w^2 x^2} 
\ee

\noindent
This large $N$ background is independent of $\lambda$. 

%\noindent
%Droplet picture comments. May have to introduce the time of flight here

%\noindent
It is useful to review here the emergence of the droplet picture when $\lambda=0$. In this case, if we let 
\cite{Avan:1991kq} $p_{\pm}\equiv \partial_x \Pi(x) / N^2 \pm \pi \phi(x)$, 
then (\ref{HEff}) has a very natural phase space 
representation as 

$$
H_{eff}^{0} = \frac{N^2}{2\pi} \int_{p_{-}}^{p_+} \int dp dx \big( \frac{p^2}{2} + \frac{x^2}{2} - \mu \big). 
$$ 

\noindent
As $N \to \infty$, the boundary of the droplet is given by $p_{\pm}^2 + x^2= 2\mu= 2 w$, since 
$p_{\pm} \to \pm \pi \phi_0 (x) = \pm \sqrt{2\mu - w^2 x^2}$.  This is in agreement with the energy of the gravity
solutions considered in \cite{Lin:2004nb}, with $x_1 \to x, x_2 \to p$\footnote{In the notation of \cite{Lin:2004nb}, 
our solution is restricted to $y\to 0$}.

When $\lambda \ne 0$, all possible $\lambda$ dependence is absorbed in the definition of $\mu$, with the result that the
large $N$ background has a similar $\lambda$ independent fermionic description.

%\noindent
%\section{Spectrum}

For the small fluctuation spectrum, one shifts 

$$
\phi(x) = \phi_0 + {1\over \sqrt{\pi} N} {\partial_x \eta };
\qquad \partial_x \Pi(x) = - \sqrt{\pi} N P (x)
$$

\noindent
to find the quadratic operator

\bea
&&H_{2}^{0}= {{1 \over 2}}
\int dx (\pi\phi_{0}) P^2(x) + \frac{1}{2} \int dx (\pi\phi_0) ({\partial_x \eta })^2 \nonumber \\  
&&+ \frac{\lambda^2}{8w^4\pi} \big( \int dx x^2 \phi_0 \big)\big( \int dx {\partial_x \eta } \big)^ 2
+ \frac{\lambda}{2w^2\pi} \big( \int dx {\partial_x \eta } \big)^ 2.
\eea

\noindent
The variable with a gravity interpretation as the angular variable
in the plane of the droplet \cite{Lin:2004nb} is the classical "time" of flight $\phi$

\be\label{phio}
{dx \over d\phi} = \pi \phi_0 ; \quad x(\phi)= - \sqrt{\frac{2}{{w}}} \cos({w}\phi) ; 
\quad \pi \phi_0 =  \sqrt{2{w} } \sin ({w} \phi);
\quad 0 \le \phi \le \frac{\pi}{{w}},
\ee

\noindent
in terms of which the quadratic Hamiltonian takes the form:

\bea\label{Quadratic}
&&H_{2}^{0}= {{1 \over 2}}
\int d\phi P^2(\phi) + {1 \over 2} \int d\phi ({\partial_\phi \eta })^2 \\
&&+ \frac{\lambda^2}{8w^4} \big( \int d\phi x^2(\phi) \phi_0^2(\phi) \big)\big( \int d\phi {\partial_{\phi} \eta } \big)^ 2
+ \frac{\lambda}{2w^2\pi} \big( \int d\phi {\partial_{\phi} \eta } \big)^ 2 . \nonumber
\eea

Except for the last two terms, one recognizes the Hamiltonian of a $1+1$ massless boson. For consistency of the time
evolution of the contraint (\ref{constraint}), we impose Dirichelet boundary conditions at the classical turning points:
%Remarkably:
% yields the spectrum in the zero impurity sector 

\be \label{EigenZero}
                \psi_j (\phi) = \sqrt{\frac{2{w}}{\pi}} \sin(j{w}\phi),  \quad j=1,2 , ...
\ee

\noindent
Since then 

\be\label{tconst}
 \int d\phi {\partial_{\phi} \eta } = 0
\ee

\noindent
the last two terms in (\ref{Quadratic}) vanish, and we obtain a $1+1$ massless boson hamiltonian with spectrum

\be\label{SpecUL}
      \epsilon_j = wj \quad, j=1,2,3,...
\ee

\noindent
again independent of $\lambda$.

%\noindent
It is not dificult to see that as a result of both (\ref{constraint}) and (\ref{tconst}) 
the collective field cubic interaction is independent of $\lambda$. 

%\noindent
In addition to (\ref{CollBef}), there is in the collective field theory an extra subleading
contribution proportional to $\int dx \partial w(x) / \partial \phi(x)$, responsible for
the one loop contribution to the energy and a tadpole interaction term {\cite{COne}}.
The additional $\lambda$ dependent contribution is proportonial to a 
total derivative of $x \phi(x)$. 

\noindent
\section{A simple diagrammatic test}

\noindent
Let us consider, for instance, the $g_{YM}^2$ contribution to the two point function

\be
\frac{1}{N^2} \left< \tr(A^2(t_2)) \tr ({A^{\dagger}}^2(t_1)) \right>, \quad t_2 > t_1
\ee

\noindent
resulting from the quartic interaction in (\ref{HamAA}). This can be done using the propagator

\be
    \left<0|T ( \tr(A_{i_1 j_1}(t_2) A^{\dagger}_{i_2 j_2}(t_1)) |0 \right> = 
\theta (t_2 - t_1) e^{-iw(t_2-t_1)} \delta{i_1 j_2} \delta{i_2 j_1}.
\ee

\noindent
Consider first the $\tr(A^{\dagger}A^{\dagger} A A)$ interaction. There are two types of 
connected planar diagrams, one the usual connected diagram without self contractions and the other
involving the ``dressing" of an internal leg. For the first, one obtains

\be\label{DiagO}
\frac{g_{YM}^2}{4 w^2} \, . 4N \, . \, \int_{t_1}^{t_2} dt e^{-i2w(t_2-t)} e^{-i2w(t-t_1)}=
\frac{\lambda}{4 w^2} \, . 4 \, . \, ({t_2}-{t_1}) e^{-i2w(t_2-t_1)}
\ee     

\noindent
For the other type of diagram, one obtains

\be\label{DiagT}
\frac{g_{YM}^2}{4 w^2} \, . 8N \, . \, e^{-iw(t_2-t_1)} \int_{t_1}^{t_2} dt e^{-iw(t_2-t)} e^{-iw(t-t_1)}=
\frac{\lambda}{4 w^2} \, . 8 \, . \, ({t_2}-{t_1}) e^{-i2w(t_2-t_1)}
\ee

\noindent
Turning now to the $\tr(A^{\dagger}A A^{\dagger} A)$ interaction, one establishes that the only type of 
planar diagram that is generated is the diagram (\ref{DiagT}) with a self-contraction associated with 
the ``dressing" of an internal leg. This diagram now has a symmetry factor of $12$, and since the interaction
has the opposite sign, it exactly cancels the sum of (\ref{DiagO}) and (\ref{DiagT}).

The arguments of the previous sections show that this generalizes to general invariant external states and to
arbitrary orders of perturbation theory.

Other families of states, that are not protected by BPS arguments but with energies independent of $g_{YM}$ 
for different diagrammatic reasons, have also been reported in the literature \cite{Agarwal:2006nv}.

\section{General D terms}
Consider now the general $g_{YM}^2$ potential

\be
- g_{YM}^2 \sum_{i<j} \tr([X_i,X_j][X_i,X_j]),
\ee 

\noindent
written in terms of complex fields $Z=X_1+iX_2$,  $\phi=X_3+iX_4$ and $\psi=X_5+iX_6$. The D terms take the form

\bea\label{Dterms}
& g_{YM}^2 &\tr  (   \frac{1}{4} [Z^{\dagger},Z][Z^{\dagger},Z] + 
                     \frac{1}{4} [\phi^{\dagger},\phi][\phi^{\dagger},\phi] + 
                     \frac{1}{4} [\psi^{\dagger},\psi][\psi^{\dagger},\psi] \\
                  && + \frac{1}{2} [\phi^{\dagger},\phi][\psi^{\dagger},\psi] +
                     \frac{1}{2} [\phi^{\dagger},\phi][Z^{\dagger},Z] +
                     \frac{1}{2} [\psi^{\dagger},\psi][Z^{\dagger},Z] ) \nonumber
\eea

\noindent
They can be equivalently written as

\be\label{Square}
 \frac{g_{YM}^2}{4} \tr \left( ([Z^{\dagger},Z] + [\phi^{\dagger},\phi] +  [\psi^{\dagger},\psi])
([Z^{\dagger},Z] + [\phi^{\dagger},\phi] +  [\psi^{\dagger},\psi]) 
\right).
\ee

\noindent
One introduces, as before, matrix valued creation and annihilation operators

\be
Z=\frac{1}{\sqrt{w}}  (A + B^{\dagger}) \,, \, \phi=\frac{1}{\sqrt{w}}  (C + D^{\dagger}) \, , \, 
\psi=\frac{1}{\sqrt{w}}  (E + F^{\dagger}).
\ee

\noindent
If one is interested only in correlators of chiral primaries such as
$\tr ({Z^{\dagger}}^{m_1} {\phi^{\dagger}}^{q_1} {\psi^{\dagger}}^{s_1}....)$, these can be
excited from the vacuum as 
$\tr ({A^{\dagger}}^{m_1} {C^{\dagger}}^{q_1} {E^{\dagger}}^{s_1}....)$ and described in 
terms of hermitean matrices $M$,$Q$ and $S$. Using an argument similar to the the general argument 
of Section $2$, and taking into account the special form of (\ref{Square}), we conclude that the terms in (\ref{Dterms})
involving only $A$, $C$ and $E$ oscillators (and their conjugates) do not contribute to these amplitudes.

\section{Conclusion}

\noindent
In this letter, the large $N$ limit of the system  

\be
\hat{H}
=
w \tr (A^{\dagger} A)  +
\frac{g_{YM}^2}{4w^2} \tr [A^{\dagger},A][A^{\dagger},A].
\ee

\noindent
when restricted to the subsector of chiral primary states $\tr({A^{\dagger}}^n)$
has been shown to be independent of $g_{YM}$. In order to explicitly confirm this result, 
the collective field theory method has been generalized to include sigma model type kinetic energy operators.
This ``non-renormalization" result applies to more general D-terms.

\noindent
\section{Acknowledgment}
I wish to thank Robert de Mello Koch for reading the manuscript and for useful comments.

\end{document}